\documentclass[11pt]{article}

\usepackage[latin1]{inputenc}
\usepackage[T1]{fontenc}
\usepackage{amsmath,amsfonts,amssymb,bm}
\usepackage{pslatex}
\usepackage{graphicx,color}
\usepackage[a4paper,left=2.5cm,right=2.5cm,top=3cm,bottom=3cm]{geometry}
\usepackage{fancyhdr}

\usepackage{multicol}

\makeatletter

\newcommand{\titlerunning}[1]{\def\@titlerunning{#1}}
\newcommand{\authorrunning}[1]{\def\@authorrunning{#1}}
\newcommand{\address}[1]{\def\@address{#1}}
\fancypagestyle{plain}
\AtBeginDocument{\lhead{\textit{\@authorrunning}}\chead{}\rhead{\textit{\@titlerunning}}}

\def\@abstract{\normalsize\textbf{Abstract. }} \def\end@abstract{\par}
\def\@keywords{\normalsize\textbf{Keywords. }} \def\end@keywords{\par}
\newtoks\abstract \newtoks\keywords

\def\acknowledgments{\@startsection{subparagraph}{6}{\z@}{-24pt plus 6pt minus 1pt}{-.5em}
{\bfseries}*{Acknowledgments. \/}}
\newcommand{\graspaXVline}{\textcolor[RGB]{255,172,89}{\rule{15.7cm}{2pt}}}

\renewcommand{\@maketitle}{
\newsavebox{\foo}
\savebox{\foo}{
\begin{minipage}[t]{15.7cm}
    \par\hspace{24.75em}

\begin{minipage}[c]{3.5cm}
\includegraphics[scale=0.2]{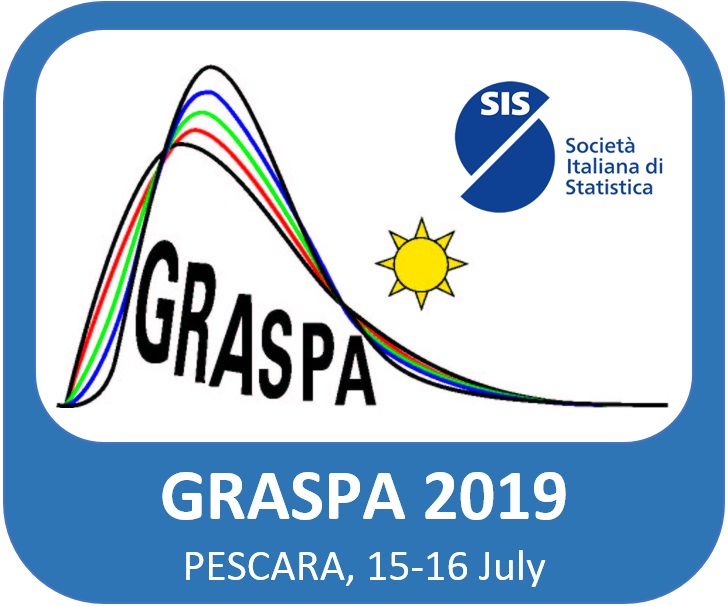}
\end{minipage}
\begin{minipage}{11.5cm}
\begin{center}
        \Large\bfseries\@title
        \par\vspace{1em}
        \normalfont\normalsize\noindent\@author
        \end{center}
\end{minipage}

    \par\vspace{0.5em}\graspaXVline\par\vspace{0.5em}
    \par\vspace{1em}\itshape\small
    \@address
    \par\vspace{1em}\noindent\rule{\linewidth}{.5pt}
    \par\vspace{1em}
    \begin{@abstract}\the\abstract\end{@abstract}
    \par\vspace{1em}
    \begin{@keywords}\the\keywords\end{@keywords}
    \par\vspace{1em}\noindent\rule{\linewidth}{.5pt}
\end{minipage}}
\par\vspace{1em}\noindent\usebox{\foo}
\vspace{1em}}
\renewenvironment{thebibliography}[1]
     {\section*{\large\refname}
      \small\@mkboth{\MakeUppercase\refname}{\MakeUppercase\refname}
      \list{\@biblabel{\@arabic\c@enumiv}}
           {\settowidth\labelwidth{\@biblabel{#1}}
            \leftmargin\labelwidth
            \advance\leftmargin\labelsep
            \@openbib@code
            \usecounter{enumiv}%
            \let\p@enumiv\@empty
            \renewcommand\theenumiv{\@arabic\c@enumiv}}
      \sloppy
      \clubpenalty4000
      \@clubpenalty \clubpenalty
      \widowpenalty4000%
      \sfcode`\.\@m}
     {\def\@noitemerr
       {\@latex@warning{Empty `thebibliography' environment}}
      \endlist}
\makeatother

\title{Multiresolution Decomposition of Areal Count Data}
\titlerunning{Multiresolution Decomposition of Areal Count Data}

\author{R. Flury$^1$ and R. Furrer$^2$}
\authorrunning{R. Flury \emph{et al.}}

\address{$^1$ Department of Mathematics, University of Zurich, Switzerland; roman.flury@math.uzh.ch\\
         $^2$ Department of Mathematics \& Department of Computational Science, University of Zurich, Switzerland; reinhard.furrer@math.uzh.ch}

\abstract{
  Multiresolution decomposition is commonly understood as a procedure to capture scale-dependent features in random signals.
  Such methods were first established for image processing and typically rely on raster or regularly gridded data.
  In this article, we extend a particular multiresolution decomposition procedure to areal count data, i.e.~discrete irregularly gridded data.
  More specifically, we incorporate in a new model concept and distributions from the so-called Besag--York--Molli\'{e} model to include a priori demographical knowledge.
  These adaptions and subsequent changes in the computation schemes are carefully outlined below, whereas the main idea of the original multiresolution decomposition remains.
  Finally, we show the extension's feasibility by applying it to oral cavity cancer counts in Germany.}

\keywords{
    Spatial scales; Lattice data; Intrinsic GMRF; Besag--York--Molli\'{e} model; MCMC.
}

\begin{document}

\maketitle
\thispagestyle{empty}


\section{Introduction}
  Decomposing an observed signal or spatial field into scale-dependent components allows recognizing its inherent and prominent features.
  Those features give insight to where local or global phenomena manifest themselves and assist in understanding the structure of hierarchical information.
  Holmstr\"om et al.~(2011) proposed a procedure in the tradition of image processing that hence is applicable to Gaussian data distributed on regular grids~\cite{H01}.
  We extend this method to count data which is potentially observed on an irregular grid, often termed \lq{areal count data}\rq~\cite{C93}.
  The original multiresolution decomposition approach can be divided into three individual steps: 1)~spatial field resampling based on a Bayesian hierarchical model, 2)~smoothing on multiple scales, then calculating differences between these smooths to specify details for each resampled field separately, and 3)~posterior credibility analysis.
  In the following paragraphs we summarize a) the Bayesian hierarchical model for step 1) and b) how to calculate differences between smooths in step 2).
  Those are the relevant parts in the procedure for the proposed extension, outlined in Section~2.
  The original multiresolution decomposition assumes that an observed field $\boldsymbol{y}$ consists of the true field $\boldsymbol{x}$ and additive noise.
  Based on these flexible model assumptions the hierarchical model is constructed.

  a) Bayesian hierarchical model: the true field $\boldsymbol{x}$ is presumed to follow a Gaussian distribution, which implies a selfsame likelihood function.
  Its positive valued variance is modeled with a scaled--inv--$\chi^2$ prior and the spatial component of the field $\boldsymbol{x}$ is captured with an intrinsic Gaussian Markov random field (IGMRF) using a precision matrix $\boldsymbol{Q}$~\cite{R05}.
  With those choices, the resulting marginal posterior is of closed form and corresponds to a multivariate t-distribution~\cite{E05}.

  b) Calculate differences between smooths: the proposed penalty smoother is defined as $\boldsymbol{S}_{\lambda} = (\mathbf{I} + \lambda\boldsymbol{Q})^{-1}$, where $\lambda$ is the scale or smoothing parameter, such that $0 = \lambda_1 < \lambda_2 < \ldots < \lambda_L = \infty$.
  The spatial field $\boldsymbol{x}$ is interpreted as random vector, $\boldsymbol{S}_{\lambda_1}\boldsymbol{x} = \boldsymbol{x}$ defines the identity mapping and $\boldsymbol{S}_{\lambda_L}\boldsymbol{x} =~\boldsymbol{S}_{\infty}\boldsymbol{x}$ the mean field.
  On the ground of those preliminaries, $\boldsymbol{x}$ can be decomposed as differences of consecutive smooths: $\boldsymbol{x} = \sum_{l=1}^{L-1} \left( \boldsymbol{S}_{\lambda_l} - \boldsymbol{S}_{\lambda_{l+1}} \right)\boldsymbol{x} + \boldsymbol{S}_{\infty}\boldsymbol{x}$.
  Scale-dependent details are then formalized as $\boldsymbol{z}_l = \left(\boldsymbol{S}_{\lambda_l} - \boldsymbol{S}_{\lambda_{l+1}} \right)\boldsymbol{x}$ for $l = 1, \ldots, L-1$ and $\boldsymbol{z}_L = \boldsymbol{S}_{\infty}\boldsymbol{x}$.

  Pivotal for a) and b) is the definition of the precision matrix~$\boldsymbol{Q}$:
\begin{equation}
    \boldsymbol{x}^\top\boldsymbol{Qx} = \sum_j \bigg( \sum\limits_{i \sim j} x_i - 4 x_j \biggr)^2,
\end{equation}
  where $i{\sim}j$ denotes neighboring grid locations.
  To ensure four neighbors at every grid location $i$, the boundary values of $\boldsymbol{x}$ are extended across the initial grid.
  This definition inherently demands the data allocated to a regular grid but bears the advantage that individual computational steps can be optimized based on $\boldsymbol{Q}$'s fast eigendecomposition, such that large dimensional problems can be solved efficiently.

\section{Extension}\label{sec:method}
  To decompose areal count data, first the resampling pattern described in a) needs modification.
  Assuming the $n$ observed counts $\boldsymbol{y}=(y_1,\dots,y_n)^\top$ are realizations from a conditionally independent Poisson distribution and the expected counts $\boldsymbol{e}=(e_1,\dots,e_n)^\top$ are known for every location in the spatial field.
  The Poisson's rate for a location $i$, is defined as the product of the expected count $e_i$ and the respective relative risk, denoted as $\exp{(\eta_i)}$.
  We construct the hierarchical model, to resample the spatial field, with the likelihood function
\begin{equation}
  \pi(\boldsymbol{y}|\eta_1,\dots,\eta_n) \propto \prod_{i=1}^{n} \exp{\bigl(y_i\eta_i - e_i\exp{(\eta_i)\bigr)}},
\end{equation}
  which corresponds to the classical Besag--York--Molli{\'e} (BYM) model~\cite{B91}.
  Whereat $\boldsymbol{\eta}$ is modeled as the composition of the true log-relative risk $\boldsymbol{u}$ and a normal zero-mean noise term $\boldsymbol{v}$, with unknown precision parameter $\kappa_{\boldsymbol{v}}$.
  Analogous to the original model, we use a first order IGMRF process to model the spatial component with accompanying precision parameter $\kappa_{\boldsymbol{u}}$, such that
\begin{equation}
  \pi(\boldsymbol{u}|\kappa_{\boldsymbol{u}}) \propto \kappa_{\boldsymbol{u}}^{\frac{n-1}{2}} \exp{\left( -\frac{\kappa_{\boldsymbol{u}}}{2} \sum_{i \sim j} (u_i - u_j)^2 \right)} = \kappa_{\boldsymbol{u}}^{\frac{n-1}{2}} \exp{\left( -\frac{\kappa_{\boldsymbol{u}}}{2} \boldsymbol{u}^\top \boldsymbol{R} \boldsymbol{u} \right)}.
\end{equation}
  Again $i{\sim}j$ denotes neighboring lattice locations but here in terms of regions sharing a common border.
  Assigning Gamma priors for both precision parameters implies a posterior distribution of non-closed form.
  Hence, we use a Gibbs sampler with a Metropolis-Hastings (MH) step to resample the log-relative risks $\boldsymbol{u}$, the noise components $\boldsymbol{v}$ and parameters~\cite{G15}.
  Finally, we exploit that the mean of a Poisson distribution is equivalent to its rate and reconstruct the spatial field with $\boldsymbol{e} \cdot \exp{(\boldsymbol{u} + \boldsymbol{v})}$, for every sampled field $\boldsymbol{u}$ and $\boldsymbol{v}$.

  We form the scale-dependent details still relying on a penalty smoother.
  Instead of using the matrix $\boldsymbol{Q}$ from the original model, we include the precision matrix $\boldsymbol{R}$ of the first order IGMRF~\cite{R05}.
  The definition of $\boldsymbol{R}$ does not limit the data to be associated with a regular grid and can be constructed based on adjacency relations of the respective observations.
  Since we use a different precision matrix, the optimized implementation relying on $\boldsymbol{Q}$ cannot be employed but we alternatively take advantage of the precision's sparse structure and apply tailored algorithms~\cite{F10}.

\section{Application}\label{sec:application}
\begin{figure}[!htb]
  \centerline{\includegraphics[scale=0.53]{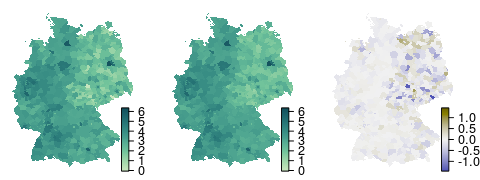}}
  \caption{Oral cavity cancer data on logarithmic scale.
           Left: the observed number of cases; middle: the mean of the reconstructed fields; right: the difference between the left and the middle panels.}
  \label{fig1}
\end{figure}

  The extension's feasibility is demonstrated on the German oral cavity cancer dataset~\cite{K00}.
  This data includes cancer counts for 544 districts of Germany over 1986--1990, as well as the expected number of cases derived demographically.
  The main bulk of the oral cavity counts range between one and hundred counts per district but single highly populated districts have up to 500.
  The data including additional relevant information is available via the \textsc{R} package \textbf{spam}~\cite{F10}.
  Following the multiresolution decomposition steps, we first resample the areal counts using suitable sampler specifications~\cite{G15} and verify the convergence of the MH sampler with common diagnostic tools~\cite{B98}.
  Figure~\ref{fig1} shows how well the reconstructed field corresponds to the original data.
  Only in northeast Germany, where the field is less smooth, the differences are larger.
  Since the BYM model was designed not to be oversensitive to extreme counts, part of the resampling difference can be explained through its damping effect~\cite{W10}.

\begin{figure}[!htb]
  \centerline{\includegraphics[scale=0.53]{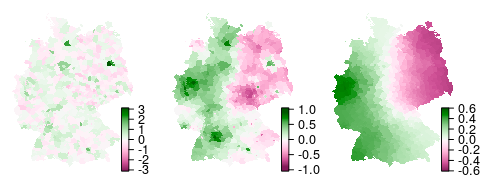}}
  \caption{Scale dependent details $\boldsymbol{z}_l=\boldsymbol{S}_{\lambda_l}{\log(\boldsymbol{e} \cdot \exp{(\boldsymbol{u} + \boldsymbol{v})} ) } - \boldsymbol{S}_{\lambda_l+1}{\log(\boldsymbol{e} \cdot \exp{(\boldsymbol{u} + \boldsymbol{v})} )}$, summarized by their posterior means.
           Left:~$\text{E}(\boldsymbol{z}_1|\boldsymbol{y})$; middle:~$\text{E}(\boldsymbol{z}_2|\boldsymbol{y})$; right:~$\text{E}(\boldsymbol{z}_3|\boldsymbol{y})$.}
  \label{fig2}
\end{figure}

\begin{figure}[!htb]
  \centerline{\includegraphics[scale=0.53]{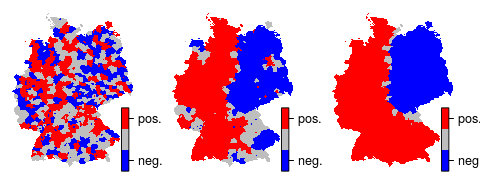}}
  \caption{Pointwise probability maps.
           Left:~$\boldsymbol{z}_1$; middle:~$\boldsymbol{z}_2$; right:~$\boldsymbol{z}_3$.
           The map indicates which features are jointly credible: blue and red areas indicate jointly credibly negative and positive areas, respectively.}
  \label{fig3}
\end{figure}

  In the second step, we choose suitable scales~(\cite{P13}) $\lambda_1 = 0$, $\lambda_2 = 1$ and $\lambda_3 = 25$ and form scale-dependent details (Figure~\ref{fig2}).
  Completing the decomposition, we calculate pointwise probability maps~\cite{H01} (Figure~\ref{fig3}).
  The detail $\boldsymbol{z}_1$ reflects spatial noise as well as the relatively low or high counts in the data.
  This is also supported by its pointwise probability map, where no large red or blue clusters are visible.
  $\boldsymbol{z}_2$ catches larger patches of districts and shows local peculiarities.
  Detail $\boldsymbol{z}_3$ consists of the largest scale range and shows the east-west or nationwide trend but this trend is less distinct compared to the more local ones, indicated by the legends of each panel.

\section{Discussion}\label{sec:discussion}
  We extended the multiresolution decomposition approach from Holmstr\"om et al. (2011), which originally processes data coming from a Gaussian distribution on a regular grid, to areal count data.
  Establishing an MH sampling model makes it possible to resample count data and use an arbitrary precision matrix.
  Employing the BYM model to include prior demographical knowledge, in the form of the known expected counts, enables us to model the data without being oversensitive to possible outliers.
  The \textsc{R} code to reproduce this example is available at https://git.math.uzh.ch/roflur/bymresa.

%
%
%

\bibliographystyle{plain}


\end{document}